\documentclass[twocolumn,superscriptaddress,showpacs,pra]{revtex4}
\usepackage{delarray}
\usepackage{amsmath, amssymb}
\usepackage{bm}
\usepackage{graphicx}
\usepackage{float}
\usepackage{placeins}
\usepackage{color}

\begin{document}
\title{A Tomographic Approach For the Early Detection of 2D Rogue Waves}

\author{Cihan Ahmet Bay\i nd\i r}
\email{cihan.bayindir@isikun.edu.tr}
\affiliation{Engineering Faculty, I\c{s}\i k University, \.{I}stanbul, Turkey}


\begin{abstract}
In this paper we propose an efficient tomographic approach for the early detection of 2D rogue waves. The method relies on the principle of detecting conical spectral features before rogue wave becomes evident in time. More specifically, the proposed method is based on constructing the 1D Radon transforms of the emerging conical 2D spectra of the wavefield using compressive sampling (CS) and then constructing 2D spectra from those projections using filtered back projection  (FBP) algorithm. For the 2D rogue wave models we use the radially symmetric Peregrine soliton and Akhmediev-Peregrine soliton solutions of the nonlinear Schr\"{o}dinger equation, which can model characteristics of the peaked structure of 2D rogue waves and their conical spectra which may be treated as a sparse signal. We show that emerging conical spectra of 2D rogue waves before they become evident in time can be acquired efficiently by the proposed method.

\end{abstract}
\maketitle

\section{Introduction}
Rogue (freak) waves are generally described as high amplitude waves with a height bigger than $2-2.2$ times the significant waveheight in a stochastic wavefield \cite{Kharif}. They have been extensively studied in recent years in the fields including but are not limited to hydrodynamics, optics, quantum mechanics, Bose-Einstein condensation, acoustics and finance, just to name a few \cite{Akhmediev2009b, bayindir2016, Akhmediev2009a, Akhmediev2011, FirstOpticalRW, Bay_Zeno, Bay_arxNoisyTun, Bay_arxChaotCurNLS}. The research has started with the investigation of the nonlinear Schr\"{o}dinger equation (NLSE). Discovery of the unexpected rational rogue wave solutions of the NLSE resulted in seminal studies of rogue waves, such as \cite{Akhmediev2009b}. Rogue wave dynamics of some of the extensions of the NLSE, such as the Sasa-Satsuma and the Kundu-Eckhaus equations, are also studied recently \cite{Soto2014RwSSchaotic, BayPRE1, BayPRE2, Bay_arxNoisyTunKEE}. It is natural to expect that in a medium whose dynamics are governed by nonlinear equations such as the NLSE and NLSE, rogue waves can also emerge, therefore investigation of the dynamics of different models needs further attention.

Development of the rogue wave early warning systems and technology is an active area of research and is crucially important for the marine environment to safeguard the ocean travel, oceanic structures and machinery such as wave energy harvesters \cite{Akhmediev2011, bayindir2016, Bay_arxEarlyDetectCS}. Two of the few early detection methods proposed in 1D are to use the emerging triangular Fourier spectra (i.e. triangular supercontinnum generation)  to detect if a rogue wave is going to emerge and to use the emerging wavelet spectra to locate its emergence location \cite{Akhmediev2011, bayindir2016, Bay_arxEarlyDetectCS}. These methods work well for the single rogue waves observed in fiber optics and hydrodynamic wave flumes and lead to early warning time scales on the order of the temporal width of the rogue wave. However enhancement of the early warning times for stochastic wavefields requires further attention and development of realistic solutions such as the development of the electronic equipment to capture rogue wave emergence may take many long efforts.

To our best knowledge the early detection rogue waves are only studied in 1D and no studies exist about the early detection mechanisms of 2D rogue waves. With this motivation, we analyze the spectral properties of 2D rogue waves. Since the correct form of the 2D NLSE is not integrable, we use a radially symmetric version of the 1D NLSE and its Peregrine and Akhmediev-Peregrine solitons solutions. Although this form of the 2D NLSE does not rely upon an analytical basis, it can exhibit the characteristics of the localized peaked structures of 2D rogue wave profiles and their conical spectral forms, very similar to the 1D case. We propose to use the emerging conical spectra of the 2D rogue waves before they become evident in time as an early detection technique and thus we discuss their dynamics. With this aim, we propose an efficient method for the acquisition of the emerging conical 2D rogue wave spectra. We first construct the 1D Radon transforms of the emerging conical 2D spectra of the wavefield using CS. Then we construct 2D spectra from those projections using FBP. Since emerging 2D conical spectra can be treated as a sparse signal, the method can successfully capture the emerging conical spectra. We numerically show that this approach can produce indistinguishable results from the classical sampling approach, but it supersedes classical sampling approach due to greatly reduced sampling requirement.

\section{Methodology}
\subsection{Review of the Nonlinear Schr\"{o}dinger Equation}
The 2D dynamics of nonlinear ocean waves, optical waves and quantum vibrations can be modeled by the 2D NLSE \cite{Akhmediev2011, Bay_Zeno, Zakharov1968}. Since 2D NLSE is not integrable, some integrable extensions are proposed in the literature which admits ration soliton solutions \cite{KunduArxiv}. However, whether they can model the realistic dynamics of 2D rogue waves or not is a question which needs further attention. In order to analyze the early detection mechanism of the 2D rogue waves, we consider radially symmetric version of the 1D NLSE in this study. Although this 2D model does not have an analytical basis it can exhibit the localized peak structures of the rogue waves. The radially symmetric rational soliton solutions of the NLSE can be used to understand the dynamics of 2D rogue waves, which are accepted as accurate rogue wave models in 1D \cite{Akhmediev2009b}. Thus we consider the radially symmetric NLSE given as
\begin{equation}
i\psi_t + \frac{1}{2} \psi_{rr} +  \left|\psi \right|^2 \psi =0
\label{eq01}
\end{equation}
where $r=\sqrt{x^2+y^2},t$ are the spatial and temporal variables, $i$ is the imaginary number and $\psi$ is the complex amplitude known as the wavefunction in optics and quantum mechanics but the wavefield envelope in hydrodynamics. This notation is mainly used in hydrodynamics and quantum mechanics whereas $t$ and $r$ axes are switched in fiber optics studies, where NLSE is used to describe the dynamics of light pulses in nonlinear fiber optical media. It is known that the NLSE given by Eq.(\ref{eq01}) admits many different types of analytical solutions among which the first and higher order rational soliton solution are considered as accurate rogue wave models \cite{Akhmediev2009b}. For stochastic wavefields where the analytical solution is unknown, the NLSE can be numerically solved by some numerical techniques such as the spectral method \cite{bay2009, demiray, BayTWMS2016,  bay_cssfm, Karjadi2010, Karjadi2012, Bay_cssfmarx, bayindir2016nature, canuto, trefethen}.  However in this study we limit ourselves with the analytical solutions of the NLSE. The radially symmetric 2D Peregrine soliton can be written as
\begin{equation}
\psi_1=\left[1-4\frac{1+2it}{1+4r^2+4t^2}  \right] \exp{[it]}
\label{eq02}
\end{equation}
where $t$ and $r$ denotes the time and space, respectively \cite{Akhmediev2009b, Peregrine}. The Peregrine soliton is only a first order rational soliton solution in the Darboux hierarchy of the NLSE and higher order rational soliton solutions do exist \cite{Akhmediev2009b}. Throughout many simulations \cite{Akhmediev2009b, Akhmediev2009a, Akhmediev2011} and some experiments \cite{Kibler}, it has been confirmed that rogue waves can be in the form of the first (Peregrine) and higher order rational soliton solutions of the NLSE. 

Second order rational soliton solution of the NLSE is Akhmediev-Peregrine soliton \cite{Akhmediev2009b}, which is considered to be a model for rogue waves with higher amplitude than the Peregrine soliton. The formula of Akhmediev-Peregrine soliton is given as
\begin{equation}
\psi_2=\left[1+\frac{G_2+it H_2}{D_2}  \right] \exp{[it]}
\label{eq03}
\end{equation}
where
\begin{equation}
G_2=\frac{3}{8}-3r^2-2r^4-9t^2-10t^4-12r^2t^2
\label{eq04}
\end{equation}
\begin{equation}
H_2=\frac{15}{4}+6r^2-4r^4-2t^2-4t^4-8r^2t^2
\label{eq05}
\end{equation}
and
\begin{equation}
\begin{split}
D_2=\frac{1}{8} & [ \frac{3}{4}+9r^2+4r^4+\frac{16}{3}r^6+33t^2+36t^4 \\
& +\frac{16}{3}t^6-24r^2t^2+16r^4t^2+16r^2t^4 ]
\label{eq06}
\end{split}
\end{equation}
where $t$ is the time and $r$ is the space parameter \cite{Akhmediev2009b}. Using Darboux transformation formalism this soliton can be obtained using the Peregrine soliton as the seed solution \cite{Akhmediev2009b}. Many numerical simulations also confirm that rogue waves in the NLSE framework can also be in the form of Akhmediev-Peregrine soliton \cite{Akhmediev2011, Akhmediev2009b, Akhmediev2009a} however to our best knowledge an experimental verification of this soliton do not exist yet. We use 2D radially symmetric versions of the Peregrine and Akhmediev-Peregrine solitons as 2D rogue wave models.

\subsection{Review of the Compressive Sampling}
\noindent Compressive sampling (CS) is an efficient sampling technique which exploits the sparsity of the signal for its reconstruction by using far fewer samples than the requirements of the classical Shannon-Nyquist sampling theorem states  \cite{Candes, Candes2006}. CS has been intensively studied as a mathematical tool in applied sciences and engineering and currently some engineering devices such as the single pixel video cameras and efficient A-D converters relies on CS algorithm.  We try to give a very brief summary of the CS in this section and refer the reader to \cite{Candes, Candes2006} for a comprehensive discussion and derivation.

Let $\psi$ be a $K$-sparse signal with $N$ elements, that is only $K$ of the $N$ elements of $\psi$ are nonzero. Using orthonormal basis transformations with transformation a matrix of ${\bf \Psi}$, $\psi$ can be represented in any transformed domain in terms of the basis functions. Most common orthogonal transformations used in the literature are the Fourier, wavelet or discrete cosine transforms. Using the orthogonal transformation it is possible to rewrite the signal as $\psi= {\bf \Psi} \widehat{ \psi}$ where $\widehat{ \psi}$ is the coefficient vector. Keeping the non-zero coefficients and discarding the zero coefficients of $\psi$, it is possible to get $\psi_s= {\bf \Psi}\widehat{ \psi}_s$  where $\psi_s$ denotes the signal with non-zero entries only.

CS algorithm guarantees that a $K$-sparse signal $\psi$ which has $N$ elements can exactly be reconstructed from $M \geq C \mu^2 ({\bf \Phi},{\bf \Psi}) K \textnormal{ log (N)}$ measurements with a very high probability. In here $C$ is a positive constant and $\mu^2 ({\bf \Phi},{\bf \Psi})$ is the mutual coherence between the sensing ${\bf \Phi}$ and transform bases ${\bf \Psi}$ \cite{Candes, Candes2006}. Taking $M$ projections randomly and using the sensing matrix ${\bf \Phi}$ the sampled signal can be written as $g={\bf \Phi} \psi$. Therefore the CS problem can be rewritten as
\begin{equation}
\textnormal{ min} \left\| \widehat{ \psi} \right\|_{l_1}   \ \ \ \  \textnormal{under constraint}  \ \ \ \ g={\bf \Phi} {\bf \Psi} \widehat{ \psi}
\label{eq07}
\end{equation}
where $\left\| \widehat{ \psi} \right\|_{l_1}=\sum_i \left| \widehat{ \psi}_i\right|$. So that, among all signals that satisfy the given constraints mentioned above, the ${l_1}$ minimization solution of the problem is $\psi_{{}_{CS}} ={\bf \Psi} \widehat{ \psi}$.  $l_1$ minimization is only one of the techniques that can be used for finding the solution of this optimization problem and other methods exist \cite{Candes, Candes2006}. Details of the CS can be seen in \cite{Candes, Candes2006}. In the current study we use the sparsity property of the 1D Radon transforms of the emerging conical 2D rogue wave spectra.

\subsection{Review of the Filtered Back Projection Algorithm}
In this section we sketch a very brief review of the FBP algorithm. The projections of a 2D function $\psi(x,y)$, which refers to the envelope of the wavefield or probability of finding an atomic particle at a specific $(x,y)$ at a given time in our study, can be computed using the Radon transform as
\begin{equation}
\widetilde{\psi}_R (r, \theta)=\int \int \psi(x,y) \delta(r-x \cos \theta-y \sin \theta)dxdy
\label{eq08}
\end{equation}
where $\theta$ is the projection angle defined from the $x$ axis. In a typical computerized tomography approach first these projections are obtained, then the full image is backprojected from these projections. However it is known that unfiltered tomographic data results in a high intensity blurring at the center of the image. In order to remove such an artifact, generally a filter is applied. Here we use a ramp filter applied in the Fourier domain as
\begin{equation}
\overline{\psi}(\rho,\theta)= F_r^{-1}  \left| k_r \right| F_r \widetilde{\psi}_R (r, \theta)
\label{eq09}
\end{equation}
where $F_r$ and $F_r^{-1}$ show the forward and inverse Fourier transform operations, respectively and $\rho$ is the radial wavenumber parameter. However, other choices of the filter also exist. Then the image can be reconstructed from these projections by means of the back projection operation given as
\begin{equation}
\psi(x,y) = B \overline{\psi}(\rho,\theta)= \int_0^\pi  \overline{\psi} (x \cos \theta+y \sin \theta) d\theta
\label{eq10}
\end{equation}
In a typical computed tomography approach, this integral is evaluated in a discrete fashion. The process summarized here is known as the FBP algorithm of the computed tomography. The reader is referred to \cite{Dudgeon}, for a comprehensive discussion of the FBP algorithm. 

\subsection{Proposed Method}

In this paper we propose using the conical spectral features before 2D rogue waves becomes evident in time as an early detection mechanism. To efficiently measure such emerging spectra we propose a tomographic approach. We first construct the 1D Radon transforms of the emerging conical 2D spectra of the wavefield using CS. This principle works because such projections are sparse signals, with nonzero entries are located around central wavenumber. Then, we construct 2D spectra from those projections using FBP. For the radially symmetric versions of the Peregrine and Akhmediev-Peregrine solitons we show that emerging conical spectral features of 2D rogue waves before they become evident in time can be acquired efficiently by the proposed method.

The tomographic method proposed in here does not necessarily have to be used with the same reconstruction techniques. For example, the CS can be utilized by random selection of the projection angles rather than equally spaced projection angles. Instead of using FBP, it is possible to use reconstruction techniques such as inverse Radon transform, Fourier domain reconstruction algorithm and ordered subsets expectation maximization techniques, just to name a few. All would have some advantages and disadvantages, but the underlying tomographic approach for the early detection of 2D rogue waves would be same in principle for all such techniques.

\section{Results and Discussion}
\subsection{Early Detection of the 2D Peregrine Soliton by the Proposed Method}
\begin{figure}[htb!]
\begin{center}
   \includegraphics[width=3.4in]{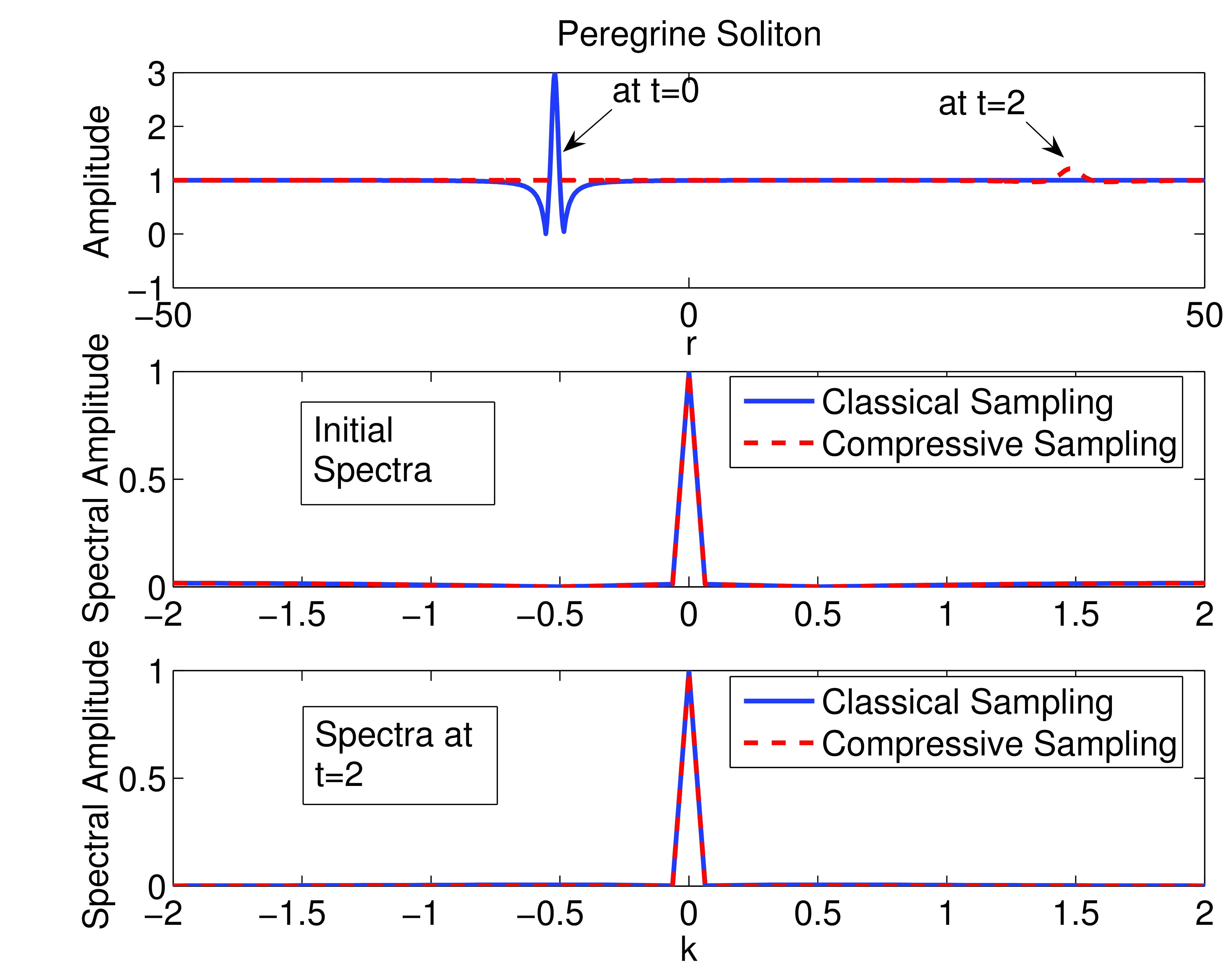}
  \end{center}
\caption{\small a) Peregrine soliton at $t=0$ and $t=2$ b) the Fourier spectrum of the Peregrine soliton at $t=0$ obtained by $N=1024$ classical and $M=64$ compressive samples c) the Fourier spectrum of the Peregrine soliton at $t=2$ obtained by $N=1024$ classical and $M=64$ compressive samples.}
  \label{fig1}
\end{figure}

In this section we numerically test the proposed algorithm for radially symmetric 2D Peregrine soliton. In the first step we take random samples along a slice in the physical domain to obtain the emerging triangular 1D spectra at various times. Then by applying the  $l_1$ minimization of the CS algorithm to those random samples acquired in the physical domain, we obtain the sparse triangular spectra. A result obtained this way is depicted in Fig.~\ref{fig1}. In Fig.~\ref{fig1}a, we show the 1D Peregrine soliton at times $t=0$ and $t=2$. Fig.~\ref{fig1}b we compare the triangular spectra of the Peregrine soliton at $t=0$ obtained by classical and compressive sampling. The normalized root-mean-square (nrms) difference between these two spectra depicted in Fig.~\ref{fig1}b are $1.56 \times 10^{-10}$. We repeat the same procedure at $t=2$ and compare the triangular spectra of the Peregrine soliton at $t=2$ obtained by classical and compressive sampling in Fig.~\ref{fig1}c, where the nrms difference between these two spectra is $7.91\times 10^{-04}$. Both of these results are obtained using $N=1024$ classical and $M=64$ compressive samples. Due to time reversal property of the phenomena studied in the frame of the NLSE, the results for $t=2$ is no different than results for the $t=-2$, thus they may be used for early detection purposes. Additionally, the detection of emerging triangular spectra can be performed starting around $t=-5$ and may be longer early detection times in the Kundu-Eckhaus equation regime \cite{BayPRE2}. We also observe that the CS is capable of constructing the triangular spectra with far fewer samples than $M=64$ when the rogue wave is at its peak at $t=0$. The use of the CS for the early detection of the 1D rogue waves is introduced and studied in \cite{Bay_arxEarlyDetectCS}.

For the 2D tomographic approach proposed above the 1D Radon transforms, i.e. the projections, of the 2D wave surface should be obtained. We obtain those projections using the perpendiculars to the slices shown above where the necessary summations are done discretely. However this is not a must, 1D Radon transforms can directly be measured using compressive samples.

\begin{figure}[htb!]
\begin{center}
   \includegraphics[width=3.4in]{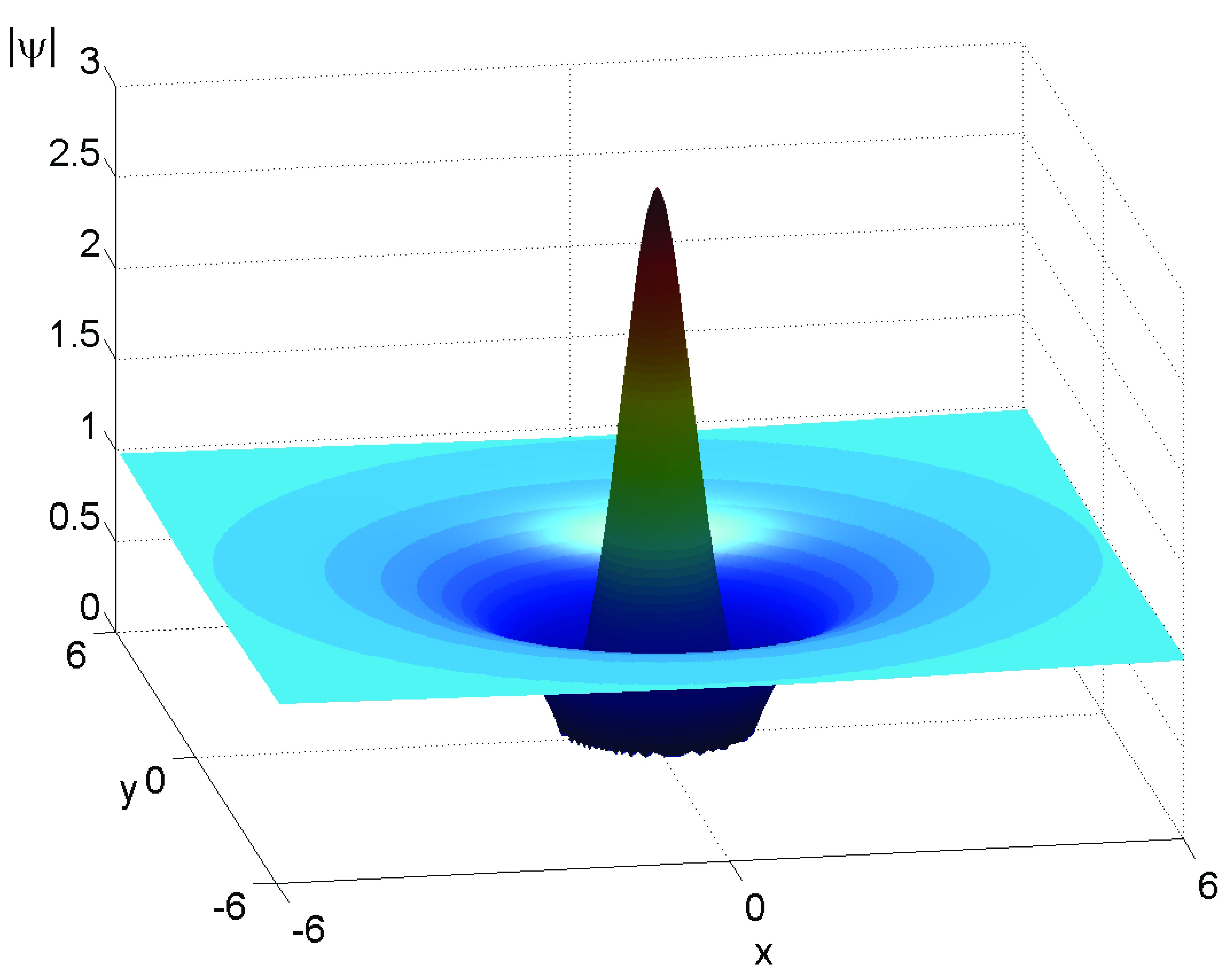}
  \end{center}
\caption{\small Radially symmetric Peregrine soliton in 2D domain at $t=0$.}
  \label{fig2}
\end{figure}

\begin{figure}[htb!]
\begin{center}
   \includegraphics[width=3.4in]{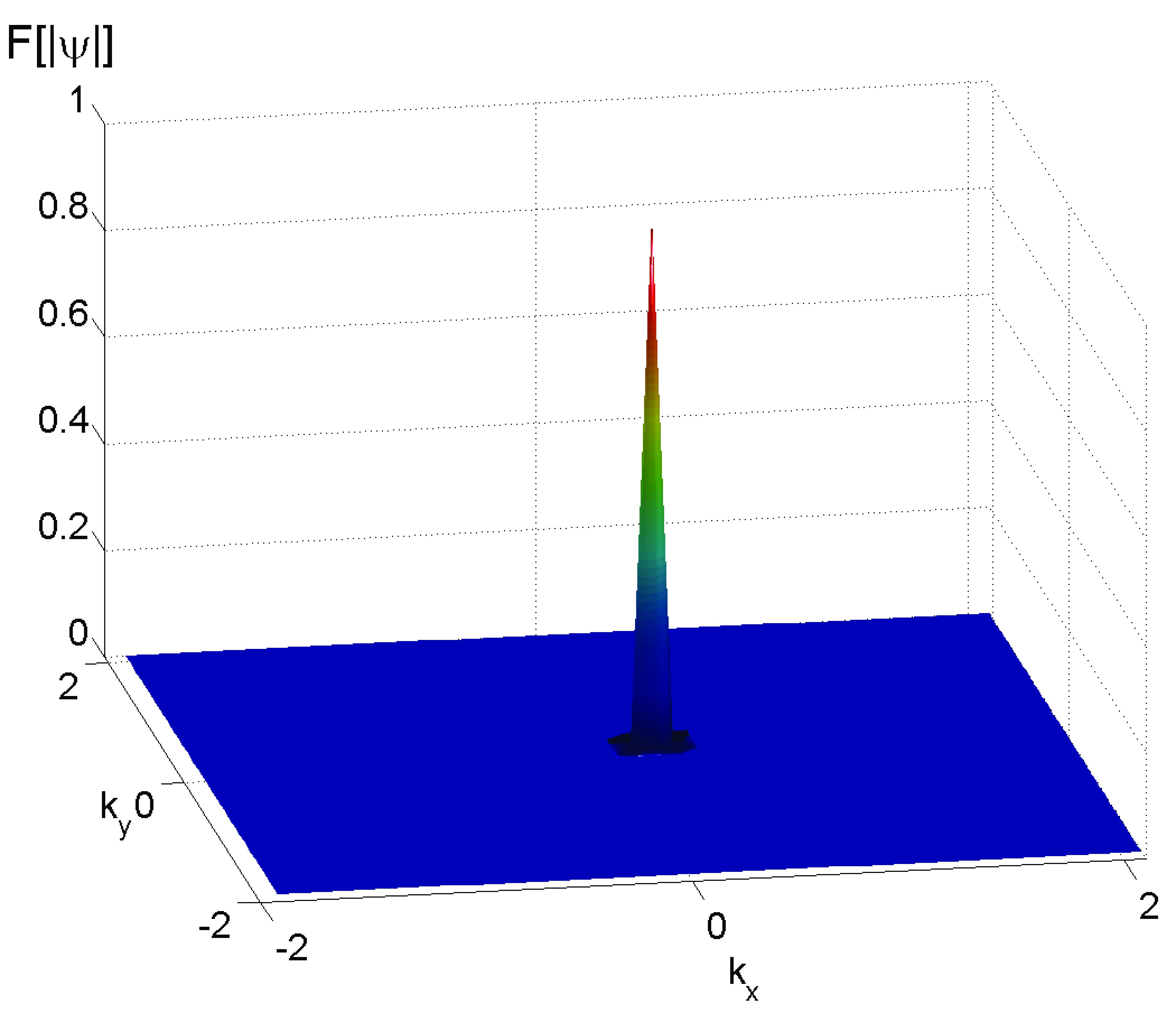}
  \end{center}
\caption{\small The Fourier spectrum of the radially symmetric Peregrine soliton at $t=0$ obtained using $N_x=N_y=1024$ classical samples.}
  \label{fig3}
\end{figure}

In Fig.~\ref{fig2} the radially symmetric 2D version of the Peregrine soliton at $t=0$ and in Fig.~\ref{fig3} its conical spectra obtained by $N_x=N_y=1024$ classical samples are depicted. This conical spectra begins to develop around $t=-5$, thus it can be used for the early detection of the 2D radially symmetric Peregrine soliton.

\begin{figure}[htb!]
\begin{center}
   \includegraphics[width=3.4in]{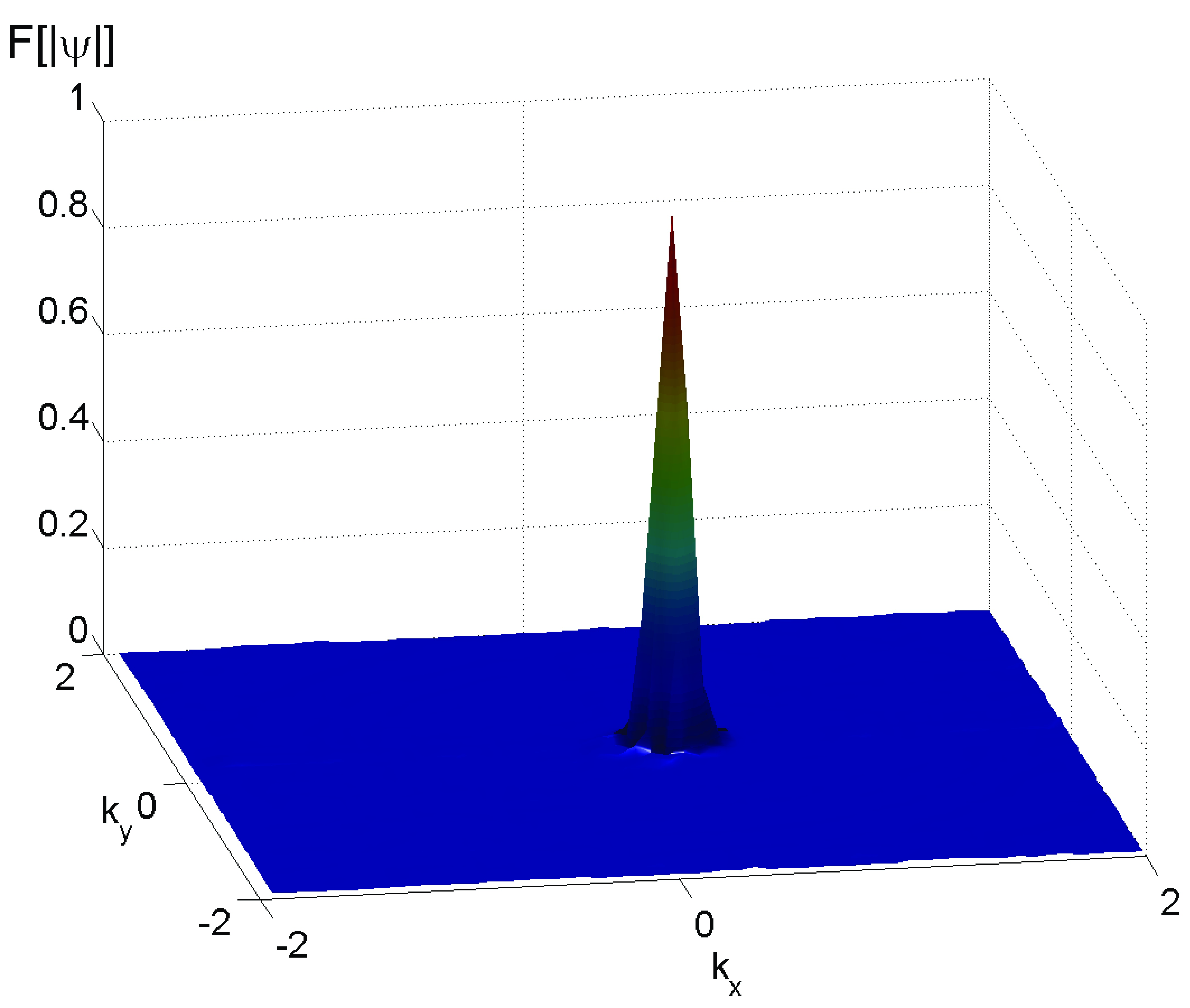}
  \end{center}
\caption{\small The Fourier spectrum of the radially symmetric Peregrine soliton at $t=0$ obtained using $M=64$ compressive samples  and FBP algorithm with angles of $0:1:179$ degrees}
  \label{fig4}
\end{figure}

In Fig.~\ref{fig4}, we present the same rogue wave spectrum obtained by the tomographic approach proposed above, where 1D Radon transforms are computed using $M=64$ compressive samples along each lines equally spaced with angles of $0:1:179$ degrees and then the FBP algorithm is used for the reconstruction of the 2D spectrum from those projections. A comparison of the results depicted in Fig.~\ref{fig3} and in Fig.~\ref{fig4} indicate that the proposed tomographic approach can successfully capture the spectral features of the 2D rogue waves, thus enables their early detection.

\begin{figure}[htb!]
\begin{center}
   \includegraphics[width=3.4in]{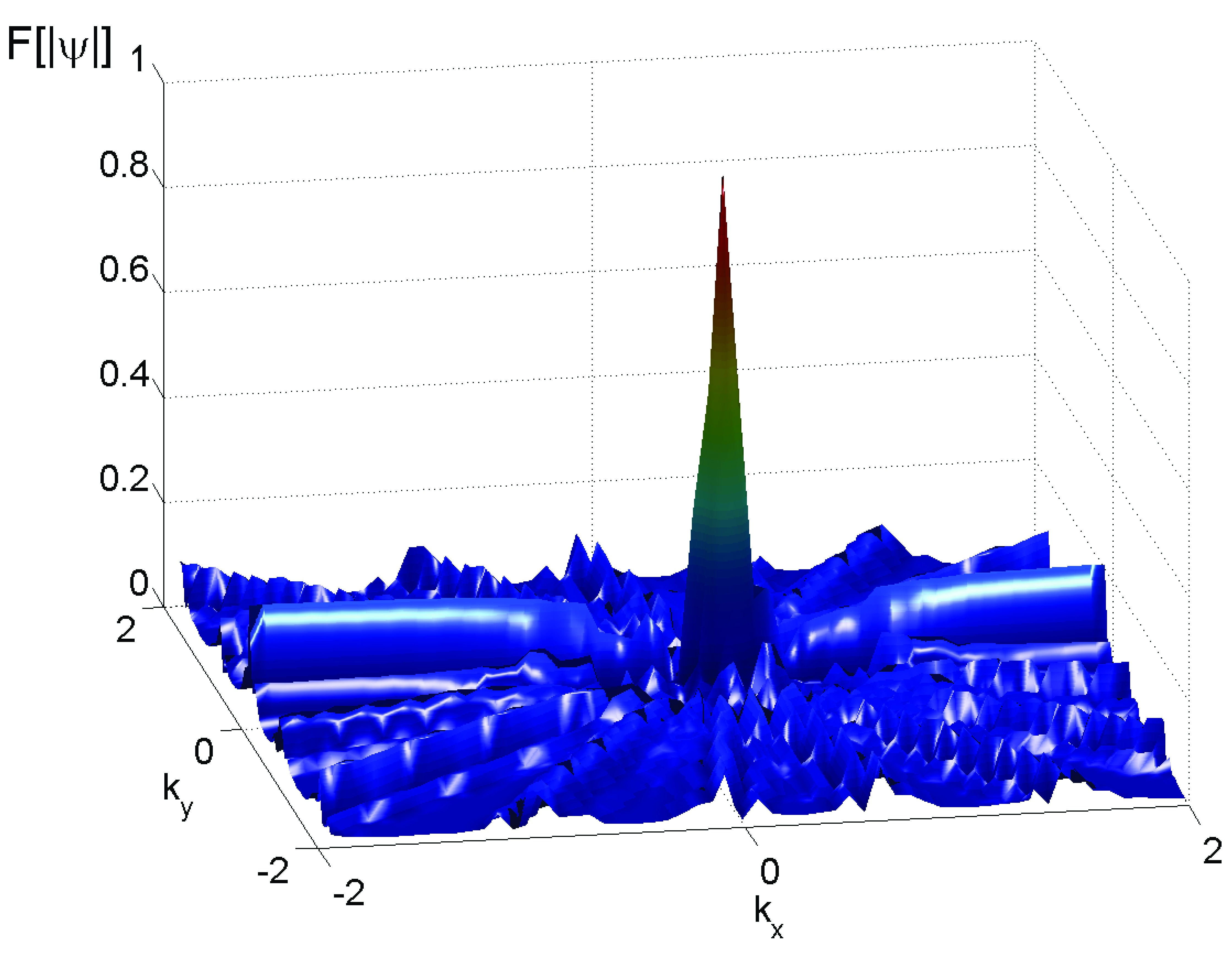}
  \end{center}
\caption{\small 3D plot of the Fourier spectrum of the radially symmetric Peregrine soliton at $t=0$ obtained using $M=64$ compressive samples and filtered backprojection algorithm with angles of $0:20:160$ degrees.}
  \label{fig5}
\end{figure}

\begin{figure}[htb!]
\begin{center}
   \includegraphics[width=3.4in]{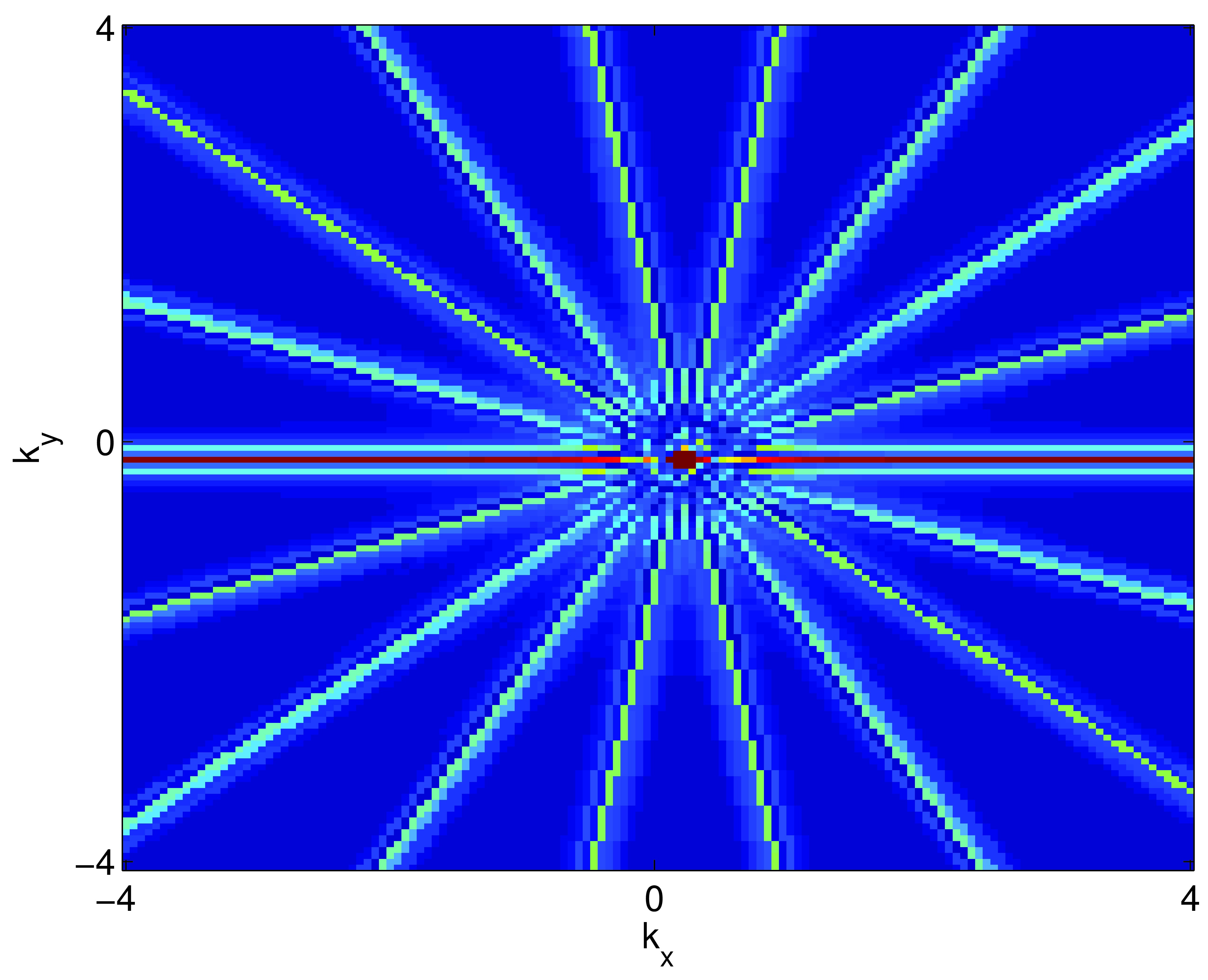}
  \end{center}
\caption{\small Contour plot of the Fourier spectrum of the radially symmetric Peregrine soliton at $t=0$ obtained using $M=64$ compressive samples on slicing lines and filtered backprojection algorithm with angles of $0:20:160$ degrees and the projections.}
  \label{fig6}
\end{figure}

In order to discuss the effects of using less projections in the tomographic approach for the early detection of the Peregrine soliton, we depict the spectra in 3D and in contour map format obtained using $9$ projection at angles of $0:20:160$ degrees in Fig.~\ref{fig5} and in Fig.~\ref{fig6}, respectively. As expected, as the number of projections decrease the capture of the conical spectral shape of the emerging rogue wave becomes harder. At central wavenumbers, the conical peak still appears and may be useful for early detection purposes, but it is surrounded by other spectral components which makes it harder to recognize if the emerging wave is a rogue wave. One possible technique to reduce the defects of small number of projections is to select projection angles randomly, which may lead to more accurate results since CS would perform better for a sparse signal when selections are random.

\subsection{Early Detection of the 2D Akhmediev-Peregrine Soliton by the Proposed Method}
Next we turn our attention to the radially symmetric Akhmediev-Peregrine soliton and assess the applicability of the proposed approach for its early detection.

\begin{figure}[htb!]
\begin{center}
   \includegraphics[width=3.4in]{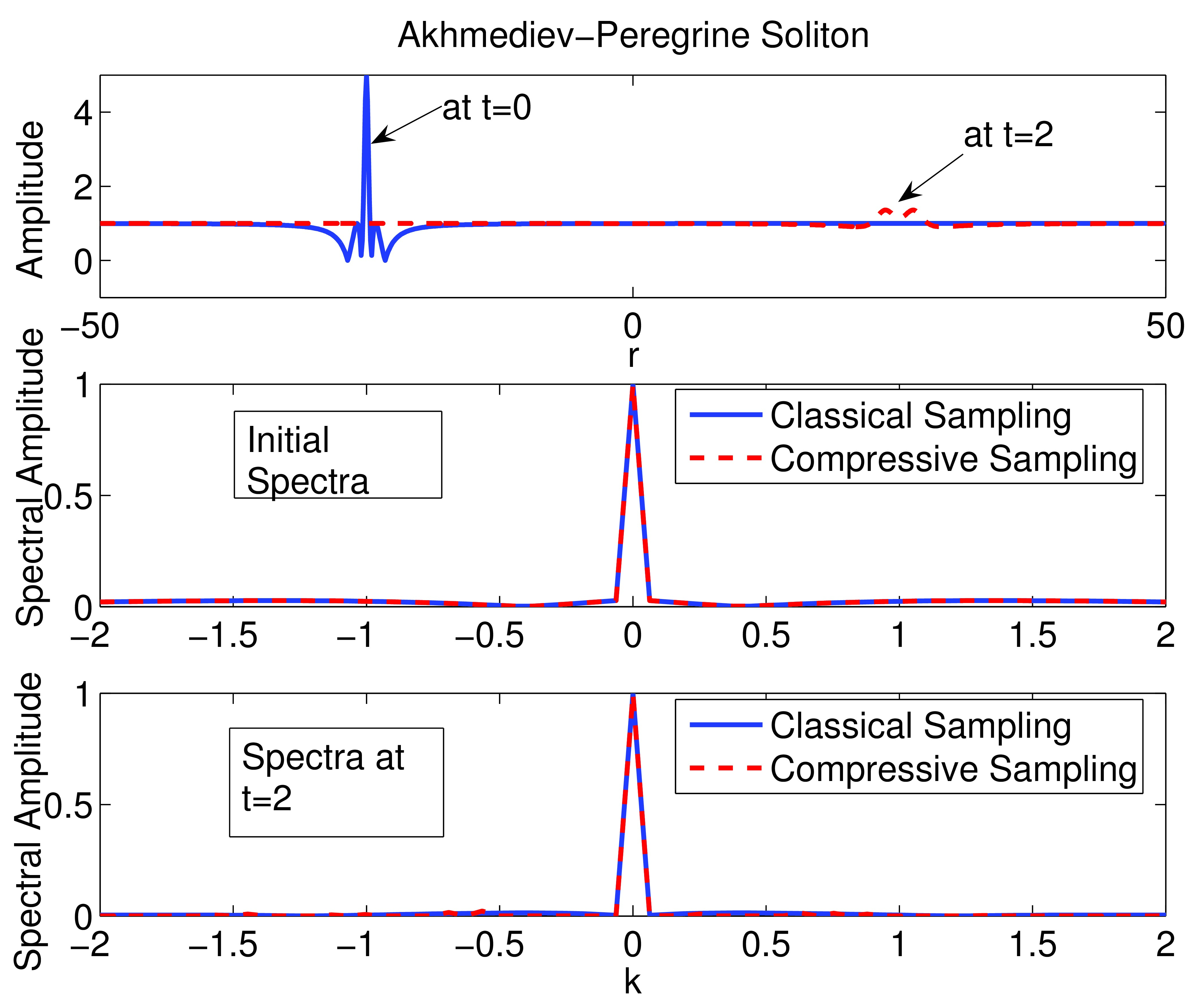}
  \end{center}
\caption{\small a) Akhmediev-Peregrine soliton at $t=0$ and $t=2$ b) the Fourier spectrum of the Akhmediev-Peregrine soliton at $t=0$ obtained using $N=1024$ classical and $M=64$ compressive samples c) the Fourier spectrum of the Akhmediev-Peregrine soliton at $t=2$ obtained using $N=1024$ classical and $M=64$ compressive samples.}
  \label{fig7}
\end{figure}
In Fig.~\ref{fig7}a, we show the 1D Akhmediev-Peregrine soliton at times $t=0$ and $t=2$. In Fig.~\ref{fig7}b we compare the triangular spectra of the Akhmediev-Peregrine soliton at $t=0$ obtained by classical and compressive sampling. The normalized root-mean-square (nrms) difference between these two spectra depicted in Fig.~\ref{fig7}b are $0.0016$. We again repeat the same procedure at $t=2$ and compare the triangular spectra of the Akhmediev-Peregrine soliton at $t=2$ obtained by classical and compressive sampling in Fig.~\ref{fig7}c, where the nrms difference between these two spectra is $0.0027$. Similar to the Peregrine soliton case, both of these results are obtained using $N=1024$ classical and $M=64$ compressive samples. We also observe that, similar to the Peregrine soliton case, the CS is capable of constructing the triangular spectra with far fewer samples than $M=64$ when the Akhmediev-Peregrine soliton is at its peak at $t=0$  \cite{Bay_arxEarlyDetectCS}.

\begin{figure}[h!]
\begin{center}
   \includegraphics[width=3.4in]{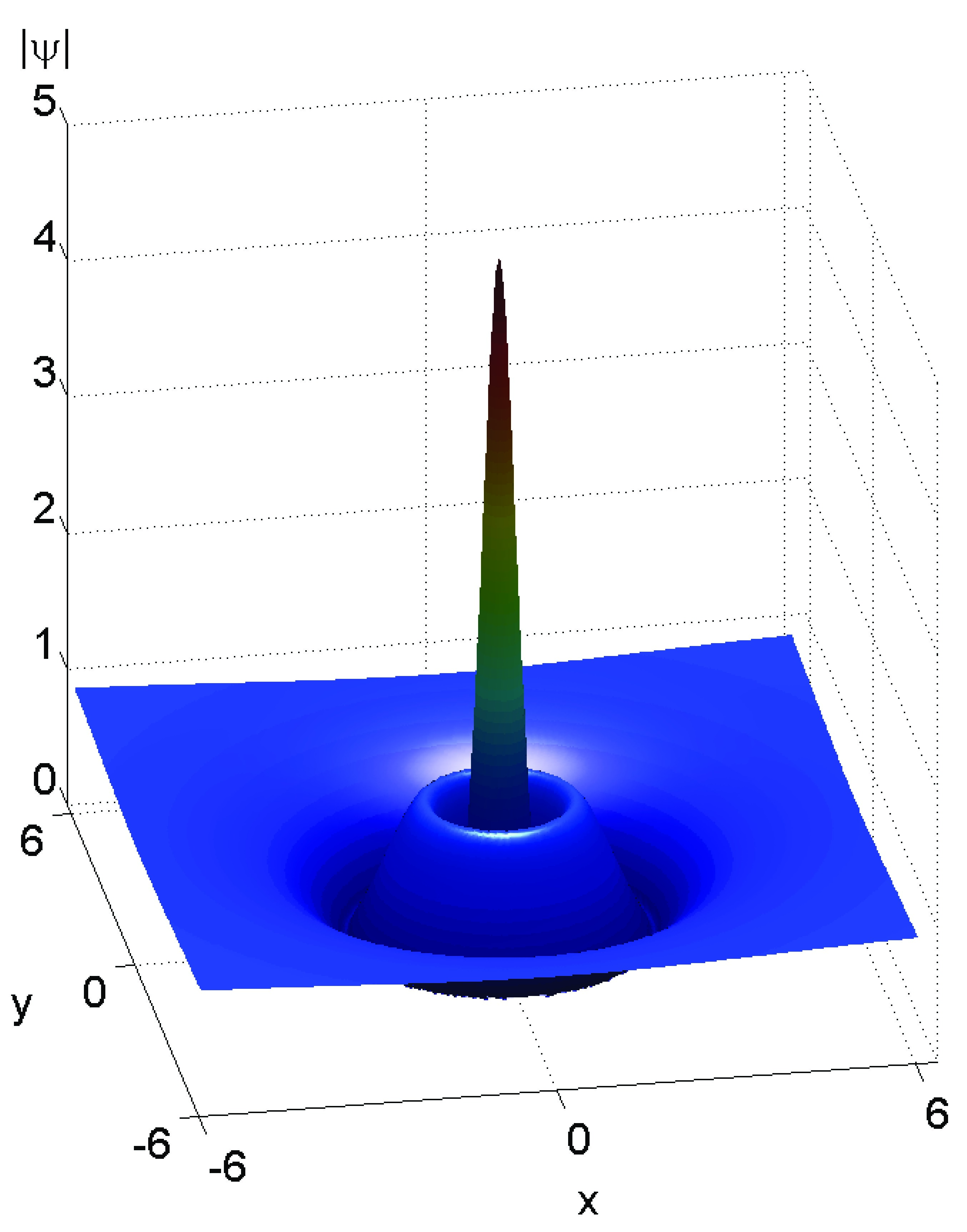}
  \end{center}
\caption{\small Radially symmetric Akhmediev-Peregrine soliton in 2D domain at $t=0$.}
  \label{fig8}
\end{figure}

\begin{figure}[h!]
\begin{center}
   \includegraphics[width=3.4in]{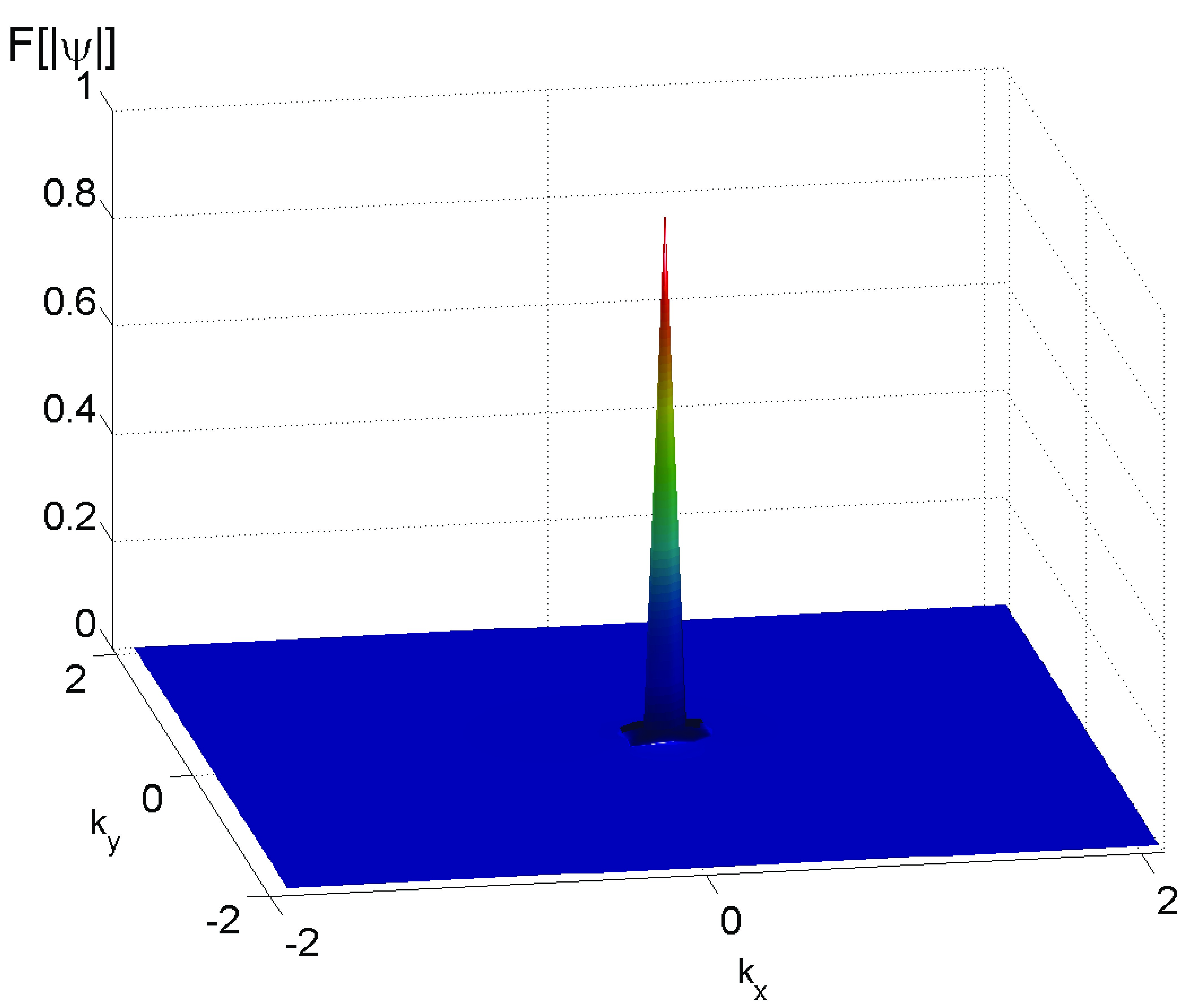}
  \end{center}
\caption{\small The Fourier spectrum of the radially symmetric Akhmediev-Peregrine soliton at $t=0$ obtained using $N_x=N_y=1024$ classical samples.}
  \label{fig9}
\end{figure}

In Fig.~\ref{fig8} the radially symmetric 2D version of the Akhmediev-Peregrine soliton at $t=0$ and in Fig.~\ref{fig9} its conical spectra obtained by $N_x=N_y=1024$ classical samples are depicted. This conical spectra begins to develop around $t=-5$, thus it can be used for the early detection of the 2D radially symmetric Akhmediev-Peregrine soliton, as in the case of the Peregrine soliton discussed above.

\begin{figure}[h!]
\begin{center}
   \includegraphics[width=3.4in]{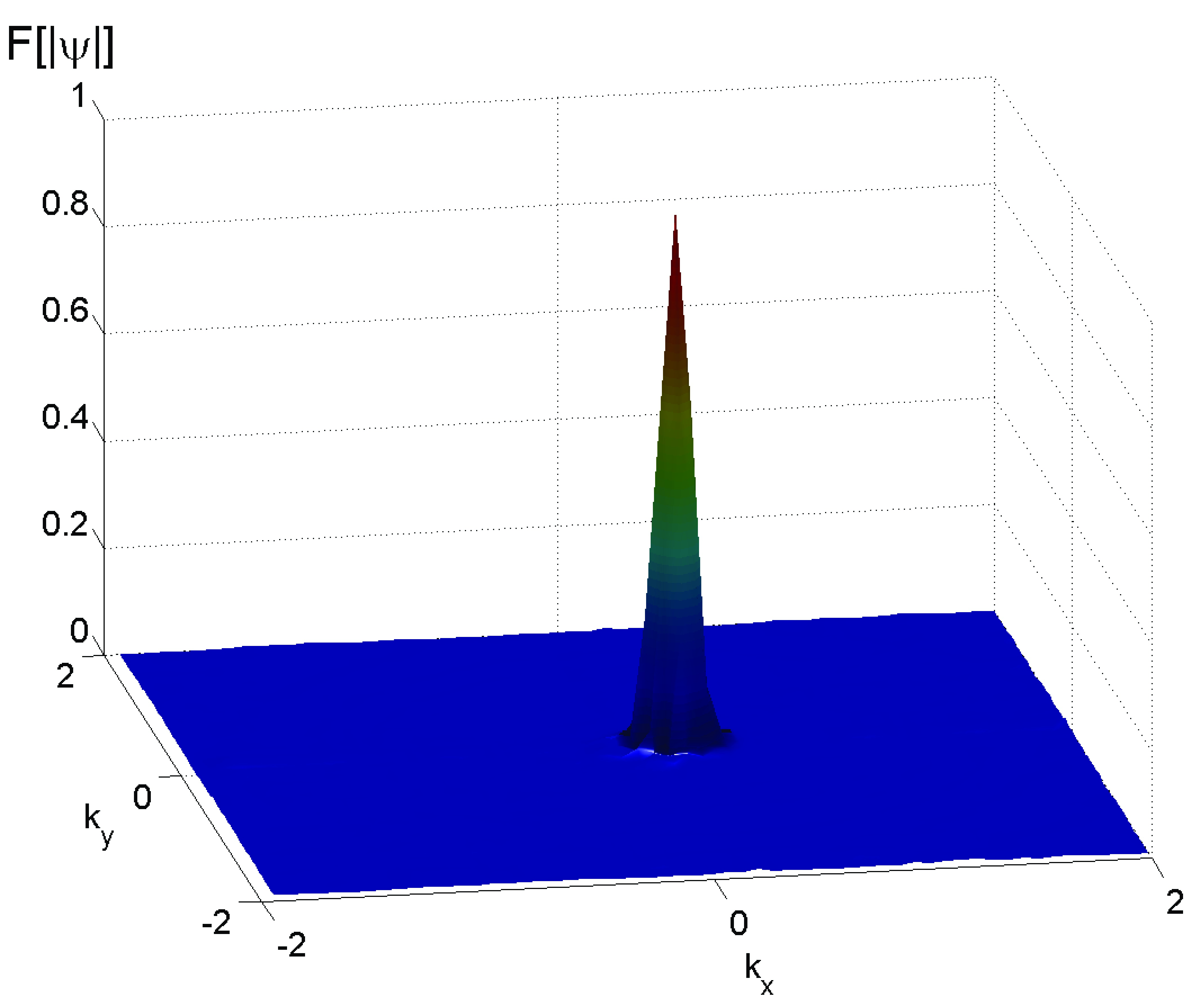}
  \end{center}
\caption{\small The Fourier spectrum of the radially symmetric Akhmediev-Peregrine soliton at $t=0$ obtained using $M=64$ compressive samples and FBP with angles of $0:1:179$ degrees}
  \label{fig10}
\end{figure}

In Fig.~\ref{fig10}, we present the same Akhmediev-Peregrine rogue wave spectrum obtained by the tomographic approach proposed above, where 1D Radon transforms are computed using $M=64$ compressive samples along each lines equally spaced with angles of $0:1:179$ degrees and then the FBP algorithm is used for the reconstruction of the 2D spectrum from those projections. Again, a comparison of the results depicted in Fig.~\ref{fig9} and in Fig.~\ref{fig10} indicates that the proposed tomographic approach can successfully capture the spectral features of the 2D Akhmediev-Peregrine soliton, thus enables their early detection before they become evident in time using spectral data.

\section{Conclusion}
In this paper we have proposed an efficient method for the early detection of 2D rogue waves. We have showed that as for the early detection of the 1D rogue waves their emerging triangular spectra can be used; so the emerging 2D conical rogue wave spectra of the 2D rogue waves can be used for their early warning. We have proposed and numerically tested a method which can efficiently be used to detect 2D rogue wave emergence. In the proposed method we have constructed the 1D Radon transforms of the emerging conical 2D spectra of the wavefield using CS and then constructed 2D spectra from those projections using FBP. We have showed that the proposed approach can successfully and efficiently detect the single rogue wave emergence in 2D, with early warning times around the temporal width of the rogue wave peak, similar to 1D case. As a future work experimental verification of the proposed method would be necessary. It should also be tested for the analytical rogue waves solutions of NLSE type equations which are physically significant, as well as for the stochastic wavefields that are triggered by the modulation instability. Additionally, other options for the tomographic acquisition technique do exist. These include but are not limited to using CS with random projections instead of equally spaced projections and using other reconstruction algorithms such as the inverse Radon transform, Fourier domain reconstruction algorithm and the ordered subsets expectation maximization techniques instead of the FBP.


\end{document}